\documentclass[aps,superscriptaddress]{revtex4}

\usepackage{amsmath,graphics}

\begin{document}

\title{A form factor approach to finite temperature correlation functions 
in $c=1$ CFT}
\author{S.~Peysson}
\email{speysson@science.uva.nl}
\affiliation{Inst. for Theoretical Physics, University of Amsterdam, 
Valckenierstraat 65, 1018 XE Amsterdam, The Netherlands}
\author{K.~Schoutens}
\email{kjs@science.uva.nl}
\affiliation{Inst. for Theoretical Physics, University of Amsterdam, 
Valckenierstraat 65, 1018 XE Amsterdam, The Netherlands}
\affiliation{Department of Physics, P.O.Box 400714, University of Virginia,
Charlottesville, VA 22904-4714, USA}

\date{May 06, 2002}

\begin{abstract}
The excitation spectrum of specific conformal field theories (CFT) with
central charge $c=1$ can be described in terms of quasi-particles with 
charges $Q=-p,+1$ and fractional statistics properties. Using the 
language of Jack polynomials, we compute form factors of the charge 
density operator in these CFTs. We study a form factor expansion for 
the finite temperature density-density correlation function, and find 
that it shows a quick convergence to the exact result. The low-temperature 
behavior is recovered from a form factor with $p+1$ particles, while the  
high-temperature limit is recovered from states containing no more 
than 3 particles.

\end{abstract}

\maketitle

\section{Introduction}

One of the extraordinary features of low-dimensional condensed matter systems 
is that they possess fractional characteristics.
Very soon after the discovery of the fractional quantum Hall effect, it was 
understood that the excitations of such states of matter carry both 
fractional charge and fractional exchange statistics.
Later, Haldane \cite{HaldanePRL67} proposed an interpretation of these 
statistics as being 
exclusion statistics, a concept that can be defined for any dimension.
It states that the presence of a particle $i$ restricts the dimension 
of the Hilbert space that is available to another particle $j$ by 
an amount $g_{ij}$, called the exclusion statistics parameter. This 
generalizes the Pauli principle for fermions.

The low-energy effective theory for a fractional quantum Hall (fqH) 
system is a chiral conformal field theory (CFT), and it has been found 
\cite{vanElburgPRB58} that these edge
theories can be analyzed in terms of quasi-particles with fractional
charge and statistics. More general chiral RCFT spectra can be 
built out of fractional statistics excitations through what is known as 
the ``universal chiral partition function'' \cite{Berkovich99}.
In the following, we will be interested in the edge CFT of a principal 
Laughlin state at filling fraction $\nu=1/p$, where $p$ is an integer.
The edge theory is that of a compactified boson at radius $R^2=p$.
The chiral Hilbert space of the theory has been understood 
\cite{vanElburgPRB58} as a collection of multi-particle states built out 
of fundamental quasi-particle excitations. For $\nu=1/p$, they are edge 
electrons, of charge $(-e)$, and edge quasi-holes, of charge $(e/p)$, 
both described by primary fields of the CFT.

It was shown \cite{SchoutensPRL79,vanElburgPRB58} that these quasi-particle 
excitations obey fractional (exclusion) statistics based on the matrix
\begin{equation}
\label{eq:gmatrix}
\boldsymbol{g}=\left(\begin{array}{cc} p & 0 \\ 0 & 1/p \end{array}\right).
\end{equation}
It means that edge electrons and edge quasi-holes are $p$- and $1/p$-ons, 
with no mutual exclusion. The $g=p$ and $g=1/p$ particles enjoy duality 
properties, in that their thermodynamic distribution functions are dual 
to each other.

These properties are reminiscent of the Calogero-Sutherland (CS) model.
Indeed, the chiral CFT for $\nu=1/p$ can be identified with the continuum 
limit of a CS model with interaction parameter equal to $p$
\cite{IsoNPB443}.
The relation between the fqH basis and the CS basis, defined in terms of 
Jack symmetric polynomials, has been worked out in \cite{vanElburgJPA33}.
The Jack polynomial technology is then available to compute the action of 
observables on multi-particle states. From there, form factors can be 
obtained, as has been shown in \cite{vanElburgJPA33}. Proceeding in this 
manner, one does not have to rely on form factor axioms for integrable 
field theories (IFT), since explicit computations in a regularized CFT 
can be performed.

It has been proposed that, in integrable field theories, 
correlation functions at finite temperature can be represented 
in a form factor expansion. In such an expansion one adds the 
contributions to the correlator coming from specific multi-particle
states, each weighted by an appropriate multi-particle distribution 
function. There is an ongoing discussion on how precisely these
ideas can be implemented for IFTs 
\cite{LeClairNPB552,DelfinoJPA34,MussardoJPA34,KonikXXX,SaleurNPB567}.
In the context of $c=1$ CFTs, a form factor expansion for a correlator
can be compared with the exact result obtained thanks to the KZ 
equations. In \cite{vanElburgJPA33}, the expansion of a specific 
electron Green function in the $\nu=1/p$ chiral CFT was shown to 
converge quickly in terms of the number of excitations considered. 
We report here on progress made in the same direction for the 
density-density correlation function, paying special attention to 
low- and high-temperature limits. At the technical level our new 
results include

\begin{description}
\item{(i)} an interpretation of the quasi-particle states with non-monotonic
    ordering of the $m_i$ and $n_j$, in terms of `vanishing' and
    `scattered states', see eq.~(\ref{eq:scatt}),
\item{(ii)} exact results for the action of the density operator on states
     with up to three particles, 
\item{(iii)} a precise definition of the irreducible part of form factors
     with more than a single quasi-hole, in which case the sector 
     index `Q' plays an important role, see eq.~(\ref{eq:defirr-qh}), 
\item{(iv)} consistency checks using the sum-rule eq.~(\ref{eq:sumrule}), 
     based on the 
     Sugawara form of the Virasoro operator in a CFT with $U(1)$ symmetry.
\end{description}

Analyzing the form factor expansion for the density-density 
correlator, we 
demonstrate that the leading asymptotic behavior for low
temperature is given by the `exciton' configuration with
a single electron and $p$ quasi-holes. More surprisingly, we
show by explicit computation that the high-temperature limit (meaning 
the $\beta\varepsilon\to 0$ limit) of the Green function 
$\langle \rho(-\varepsilon)\rho(\varepsilon)\rangle_T$, is recovered from 
form factors containing
no more than 3 particles. This remarkable result is established
with the help of the identity eq.~(\ref{eq:ng-identity}) satisfied by 
the thermodynamic distribution function $\bar{n}_g(\varepsilon)$ for particles
satisfying exclusion statistics with parameter $g$.

Our computations in this paper can be viewed as a dry run for 
the computation of non-trivial transport computations in
systems that are not CFTs but that do have quasi-particles
with well-defined exclusion statistics properties (examples
are finite-$N$ CS models or Haldane-Shastry spin chains). 
For the formalism to become of practical use, it needs
further streamlining, possibly by making and exploiting
a connection with the axiomatic approach to form factors,
based on scattering data in IFT.

This paper is organized as follows.
In section \ref{sec:basis}, we introduce our notations by recalling the fqH basis construction in \cite{vanElburgJPA33}.
We give the basic properties of the excitations and their expressions in terms of Jack polynomials.
In section \ref{sec:zoo}, we give the expressions for form factors up to 3 particles for the density operator.
In section \ref{sec:expansion} we turn to the form factor expansion for the 
density-density correlation function.
Using the results of the preceding section, we show what are the different 
contributions to the low-temperature and high-temperature limits.
Exact numerical results will sustain our arguments in favor of a quick 
convergence for the full correlation function.

\section{Fractional quantum Hall systems and Jack polynomials}
\label{sec:basis}

\subsection{fqH basis}
The effective low-energy theory for a $\nu=1/p$ fqH system is a compactified boson with radius $R^2=p$, which is a chiral $c=1$ CFT.
The Hilbert space for this theory is obtained as a collection of sectors of zero modes. The sectors are labeled by an integer $Q$, which is the $U(1)$ 
electric charge measured  in units of $e/p$. The partition function is
\begin{equation}
Z={\rm Tr}\left( q^{L_0} \right) =
\sum_{Q=-\infty}^{\infty} \frac{q^{Q^2/2p}}{(q)_\infty}
\end{equation}
where $(q)_N=\prod_{l=1}^{N} (1-q^l)$.

This Hilbert space can be understood as a collection of multi-particle states, the fundamental quasi-particles being the edge electron and the edge quasi-hole, of charge $Q=-p$ and $Q=1$ respectively.
They are described by the conformal primary fields
\begin{equation}
J^{(-p)}(z)=\sum_t J_{-t} z^{t-p/2} \qquad 
\phi^+(z)=\sum_s \phi_{-s} z^{s-1/2p} \ .
\end{equation}
The Fourier modes of these operators can be interpreted as creation operators. The independent multi-particle states that generate the chiral Hilbert space were identified to be
\begin{align}
\label{eq:basis}
&|m_M,\ldots m_1;n_N\ldots n_1>^Q \equiv J_{-(2M-1)p/2+Q-m_M} \ldots J_{-p/2+Q-m_1} \,  \phi_{-(2N-1)/2p-Q/p-n_N} \ldots \phi_{-1/2p-Q/p-n1} |Q>,\\
& \quad m_M \geq \ldots \geq m_1 \geq 0 \qquad n_N \geq \ldots \geq n_1 \geq 0 \qquad (n_1>0 \text{  if  } Q<0),
\end{align}
where $|Q>$ ($Q=-(p-1),\ldots,-1,0$) is the lowest-energy state of charge $Q$. The identification was proven through the equality of the partition functions.

\subsection{Fractional statistics and duality}
Fractional exclusion statistics is a tool introduced by Haldane \cite{HaldanePRL67} for the analysis of strongly correlated many-body systems.
It is only based on the assumption that the Hilbert space is finite-dimensional and extensive, i.e. particles are excitations of the considered condensed matter system, so it is a very generic concept.
The statistics are encoded in a matrix $\boldsymbol{g}=(g_{ij})$ corresponding to the reduction of the available Hilbert space for particle of type $i$ by filling a one-particle state by a particle of type $j$.
This is then a generalization of the Pauli principle.

The thermodynamics for a gas of such particles have been worked out by Isakov, Ouvry and Wu(IOW) \cite{DasnieresDeVeigyPRL72,IsakovMPLB8,WuPRL73}.
The one-particle grand canonical partition functions $\lambda_i$ and distribution functions $\bar{n}_i$ are respectively given by the IOW equations
\begin{align}
&\left( \frac{\lambda_i-1}{\lambda_i} \right) \prod_j \lambda_j^{g_{ij}} = e^{\beta(\mu_i-\varepsilon)}\equiv z_i,\\
&\bar{n}_i(\varepsilon) = z_i \frac{\partial}{\partial z_i} \log \prod_j \lambda_j.
\end{align}

In the case of a fqH at $\nu=1/p$ the statistical matrix is given by (\ref{eq:gmatrix}), where electrons and quasi-holes are non-exclusive to each other.
For $p=1$ we recover the Fermi-Dirac distribution functions, and for $p=2$:
\begin{equation}
\bar{n}_2(\varepsilon)=\frac{1}{2} \left(1-\frac{1}{\sqrt{1+4e^{-\beta(\varepsilon-\mu)}}} \right) \qquad \bar{n}_{1/2} (\varepsilon) = \frac{2}{\sqrt{1+4e^{2\beta(\varepsilon-\mu)}}}.
\end{equation}
In general for $g$-ons, the distribution has a limiting value for 
$\varepsilon \to -\infty$ equal to $\bar{n}_g^{\text{max}}=1/g$ and 
an asymptotic behaviour for $\varepsilon\to\infty$ equal to 
$e^{-\beta\varepsilon}$.

The transport properties of $g$-ons have been worked out in \cite{IsakovPRL83}. In the fqH case, one furthermore has a duality between $g$- and $1/g$-ons, appearing in the identity
\begin{equation}
g\bar{n}_g(\varepsilon)=1-\bar{n}_{1/g} (-\varepsilon/g)/g \ ,
\label{eq:duality}
\end{equation}
related to the fact that the removal of a $g$-on corresponds to the 
creation of $g$ $1/g$-ons. This agrees with the physics of edge-to-edge 
tunneling in the 
fqH, in which the duality couples weak and strong backscattering.
It also means the quasi-particle basis proposed in (\ref{eq:basis}) is 
not unique.
Iso \cite{IsoNPB443} proposed a basis made out of particles of one kind only,
filling up 1-particle states with energies extending over both positive and 
negative values. One shall see that our 'excitation' picture is more 
practical to compute physical quantities.

The duality (\ref{eq:duality}) can be used in the evaluation of 
thermodynamic quantities.
In each $Q$ sector, the partition function decomposes in a product of the 
partition function for electrons and that for quasi-holes.
Then, the specific heat (or central charge) can be written as a sum over the 
electron and the quasi-hole contribution, and using (\ref{eq:duality}) one
finds $c=1$ independent of $p$. Depending on the sign of an imposed
Voltage, the Hall conductance is given by an electron or by a quasi-hole 
expression, of the general form
\begin{equation}
G=\bar{n}_g^{\text{max}}\frac{q^2}{h} = \frac{1}{p} \frac{e^2}{h} ,
\end{equation}
where $q=e/p$ or $q=-e$ is the charge of the quasi-particle that carries 
the current. We shall see that the duality is also instrumental in the 
computation of form factors and the evaluation of the form factor expansion.

\subsection{Correspondence with Jack polynomials}
As for the fqH multi-particle basis, it has been shown to be in one-to-one correspondence with the orthogonal eigenbasis of the Calogero-Sutherland (CS) model
\begin{equation}
H_{CS}=-\sum_{i=1}^N \frac{\partial^2}{\partial x_i^2} + \left(\frac{2\pi}{L}\right)^2 \sum_{i<j} \frac{2\lambda(\lambda-1)}{\sin^2(\pi x_{ij}/L)},
\end{equation}
in the thermodynamic limit $N\to\infty$ and with interaction parameter $\lambda=p$.
This Hamiltonian is best specified using a scalar field $\varphi(z)$ with $\partial\varphi(z)=\sum_n a_n z^{-n-1}$.
In terms of this field, the CS Hamiltonian takes the form 
\begin{equation}
H_{CS}=\frac{p-1}{p} \sum_{l=0}^{\infty} (l+1)(i\sqrt{p}a_{-l-1})
(i\sqrt{p}a_{l+1}) \, +\frac{1}{3p} \left[ (i\sqrt{p} \partial \varphi)^3 \right]_0.
\end{equation}
It is then possible to build all the eigenstates of the Hamiltonian using multi-$J$ and -$\phi$ quanta.
It is found that the states (\ref{eq:basis}) are not $H_{CS}$ eigenstates, but that they rather act as head states that need to be supplemented by a tail of subleading states.
The eigenbasis will then be denoted by
\begin{equation}
|\{m_i;n_j\}>^Q = |m_i;n_j>^Q + \ldots
\label{eq:fqHstates}
\end{equation}
We refer to \cite{vanElburgJPA33} for further details.

From a different perspective, the analysis in \cite{IsoNPB443} 
has led to a basis of eigenstates of $H_{CS}$,  specified as
\begin{equation}
|\{\mu\},q> = J^{1/p}_{\{\mu'\}}(\{p_n=\sqrt{p}a_{-n}\})|q> 
\equiv J^{1/p}_{\{\mu'\}}|q>,
\label{eq:CSstates}
\end{equation}
where the $U(1)$ charge $q$ runs over all integers, and $\{\mu\}$ runs over all Young tableaus.
$J^{1/p}_{\{\mu'\}}$ is called a Jack symmetric polynomial. They form a basis 
of the ring of symmetric functions with a given scalar product. Useful 
details about them are reported in the appendix.

A one-to-one correspondence between fqH states 
(\ref{eq:fqHstates})
and CS states 
(\ref{eq:CSstates})
is obtained through the identification \cite{vanElburgJPA33}
\begin{equation}
|\{m_i;n_j\}>^Q=|(\{m\}+N^M)\cup\{n'\},Q+N-pM>,
\label{eq:compositeYoung}
\end{equation}
with $\{m\}$ the Young tableau built with the $m_i$ quanta and $\{n'\}$ the 
dual Young tableau built with the $n_j$ quanta.

Our strategy will be the following. We will set up the form factor 
expansion in terms of the fqH basis, but do the actual computation 
of the form factors using the representation in terms of Jack 
polynomials, allowing us to exploit what is known about them. 
In the appendix we present some `Jack polynomial technology' that
we have used for computing the various form factors.
In the next section we give an overview of the form factors 
that we obtained.

\section{Overview  of form factors}
\label{sec:zoo}

In ref.~\cite{vanElburgJPA33}, form factors for the electron creation 
operator have been studied.
Here we focus on the computation of form factors for the charge density 
operator in the fqH edge at $\nu=1/p$.
Its Fourier modes are $i\rho_m=i\sqrt{p}a_m=ip_{-m}$, so that it has a nice 
interpretation in the language of Jack polynomials as a power sum.
The form factors we are searching for are
\begin{equation}
<\{\nu\},q|p_{-m}|\{\mu\},q> \ .
\label{eq:generalff}
\end{equation}
We shall obtain them by developing the product of a power sum and a Jack
polynomial on the basis of Jack polynomials.
In the former works on the Calogero-Sutherland model \cite{LesageNPB435}, 
where only zero-temperature properties were computed, one only considered
form factors with either the in or the out state equal to the vacuum.
For such cases, only the expansion of power sums on a basis of Jack 
polynomials is needed. In the case of form factors of the general form
\ref{eq:generalff}, which appear in the form factor expansion for
finite temperature correlators, it is necessary to use a less traditional 
approach, combining the expansion of the power sums in elementary symmetric 
functions and the Pieri formula written in the appendix.
This allows in principle to compute \emph{any} form factor.
We have obtained closed, analytic expressions for a number of form
factors with up to three particles. To obtain them, we have proceeded by
evaluating a few simple examples (the smaller $m$'s), conjecturing a 
general form, and then checking the conjectured form by using the 
following sum rule
\begin{equation}
\label{eq:sumrule}
\sum_{m \geq 1} {}_N<\{m_i;n_j\}| p_m p_{-m} |\{m_i;n_j\}>_N = p |\mu| \ .
\end{equation}
The sum rule follows from the so-called Sugawara form of the 
Virasoro operator $L_0$ 
\footnote{The sum rule (\ref{eq:sumrule}) can also be obtained using only Jack polynomial technology.},
\begin{equation}
L_0 = \frac{1}{2p} \rho_0^2 + \frac{1}{p} \sum_{m \geq 1} \rho_{-m}\rho_m \ .
\end{equation}
The sum rule is given in terms of normalized basis states
\begin{equation}
|\{m_i;n_j\}>_N = N_{\{m_i;n_j\}}^{-\frac{1}{2}} |\{m_i;n_j\}> \ ,
\end{equation}
with
\begin{equation}
N_{\{m_i;n_j\}} = \langle \{m_i;n_j\} | \{m_i;n_j\}> 
= j_{\mu^\prime}^{\frac{1}{p}} \ ,
\end{equation}
with $\{\mu^\prime\}$ the dual to the composite Young tableau 
defined in eq.~(\ref{eq:compositeYoung}) and $j$ the inner product 
given in eq.~(\ref{eq:Jackinner}).

We report in the following results for one- to three-particle 
states, in any $Q$-sector. In each case we give expressions for
general $p$, and then specify to the case $p=2$, which is the case 
analyzed numerically in section \ref{sec:expansion}.

\subsection{1 quasi-particle}
One finds
\begin{equation}
\label{eq:ff1p}
p_{-m} |m_1>= \pm \sqrt{pg} |m_1-m>,
\end{equation}
which for $p=2$ gives
\begin{align}
p_{-m} |\{m_1\}> &= -2 |\{m_1-m\}> \qquad \text{for electrons},\\
p_{-m} |\{n_1\}> &= |\{n_1-m\}> \qquad \text{for quasi-holes}.
\end{align}

\subsection{2 quasi-particles}
At the level of two particles, many-body effects start to appear.
It means that the density operator is not acting on each particle individually, but rather on both at the same time.
\subsubsection{2 electrons or 2 quasi-holes}
We obtain the following action
\begin{align}
\label{eq:ff2p}
&p_{-m} |\{m_2,m_1\}> = \pm m \,\sqrt{pg}  \sum_{l=0}^{m} G(m_1,m_2;m_1-l,m_2-m+l) |\{m_2-m+l,m_1-l\}>,\\ \nonumber
&G(m_1,m_2;m_1'=m_1-l,m_2'=m_2-m+l)= G_{12} \\ \label{eq:Gfunction}
 &= \sum_{i=0}^{l} \frac{i(-)^i}{l(m-l+i)} \left(\begin{array}{c} m-l+i \\ i \end{array} \right) \left(\begin{array}{c} l \\ i \end{array} \right) \frac{\Gamma(g+i)\Gamma(m_2-m_1+g+1)\Gamma(m_2'-m_1'+g-i)}{\Gamma(g-i)\Gamma(m_2-m_1+g+i+1)\Gamma(m_2'-m_1'+g)} .
\end{align}
For the expression eq.~(\ref{eq:ff2p}) to make sense, we have to 
specify the meaning of 2-particle states $|\{m_2,m_1\}\rangle$ with
$m_2<m_1$. For the case of electrons, the prescription is
\begin{eqnarray}
\label{eq:scatt}
|\{m_2',m_1'\}> 
& \Rightarrow & (-)^p \frac{N_{\{m_2',m_1'\}}}{N_{\{m_1'-p,m_2'+p\}}} 
|\{m_1'-p,m_2'+p\}>  \qquad \text{ if } m_2'\leq m_1'-2p \nonumber
\\
& \Rightarrow & 0 \qquad
  \text{ if } m_2'-m_1'= -1, -2, \ldots, -2p+1 \ .
\end{eqnarray}
This rule can be extended to multi-particle states, and to quasi-holes. 
The latter behave slightly differently: one needs at least $(p+1)$
quasi-holes to have scattered states. Clearly, the prefactor 
$\frac{N_{\{m_2',m_1'\}}}{N_{\{m_1'-p,m_2'+p\}}}$
is analogous to the scattering phase that one expects in the 
continuum limit of this theory. 

In the case of $p=2$, we have
\begin{multline}
p_{-m} |\{m_2,m_1\}> = -2 \bigg( |\{m_2-m,m_1\}> + \frac{(m_2-m_1+1)(m_2-m_1+m+3)}{(m_2-m_1+3)(m_2-m_1+m+1)} |\{m_2,m_1-m\}> \\ - 2m \sum_{l=1}^{m-1} \frac{1}{(m_2-m_1+3)(m_2-m_1+2l-m+1)}  |\{m_2-m+l,m_1-l\}> \bigg)
\end{multline}
for electrons, and
\begin{multline}
p_{-m} |\{n_2,n_1\}> = |\{n_2-m,n_1\}> \\ 
+ \frac{\Gamma(n_2-n_1+1/2)\Gamma(n_2-n_1+3/2)\Gamma^2(n_2-n_1+m+1)}{\Gamma(n_2-n_1+m+1/2)\Gamma(n_2-n_1+m+3/2)\Gamma^2(n_2-n_1+1)}) |\{n_2,n_1-m\}> \\
+ \sum_{l=1}^{m-1}\bigg(\sum_{i=0}^{l} \frac{mi}{l(m-l+i)}(-)^i \left(\begin{array}{c} m-l+i \\ i \end{array} \right) \left(\begin{array}{c} l \\ i \end{array} \right) \frac{\Gamma^2(i+1/2)}{\Gamma^2(1/2)} \\ 
\times
\frac{\Gamma(n_2-n_1+3/2)\Gamma(n_2-n_1+2l-m-i+1/2)}{\Gamma(n_2-n_1+i+3/2)\Gamma(n_2-n_1+2l-m+1/2)} \bigg) |\{n_2-m+l,n_1-l\}>
\end{multline}
for quasi-holes.
The factor appearing in front of the second term in the right hand side of 
these expressions corresponds to the reordering of the particles during the action of the density operator. We shall call it a scattering factor, though it doesn't have the same physical origin as the one appearing in (\ref{eq:scatt}).

\subsubsection{1 electron and 1 quasi-hole}
\begin{multline}
\label{eq:ffmix}
p_{-m} |\{m_1;n_1\}> =  -p |\{m_1-m;n_1\}> +\frac{(m_1+p(n_1-m)+1)(m_1+p(n_1+1))}{(m_1+p(n_1-m+1))(m_1+pn_1+1)} |\{m_1;n_1-m\}> \\
- mp(p-1) \sum_{l=1}^{m-1} \frac{1}{(m_1-m+l+p(n_1-l+1))(m_1+1+pn_1)} |\{m_1-m+l;n_1-l\}>,
\end{multline}
along with the replacement
\begin{equation}
\label{eq:replacement}
|\{-1;0\}>^Q \equiv \left\lbrace \begin{array}{c} |\{-;-\}>^{-(p-1)} \text{ for } Q=0 \\ |\{0;-\}>^{Q+1} \text{ for } Q<0 \ . \end{array} \right. 
\end{equation}

A $G$-function can also be defined here, defining it as the factor in the 
last line of (\ref{eq:ffmix}) divided by $m$.

\subsection{3 quasi-particles}

For the case of form-factors with three quasi-particles, we do not
have complete results, but we report the following. We distinguish 
between terms where a single one, or two, or all three quasi-particles 
are affected by the action of the density operator
\begin{itemize}
\item when the density operator is affecting 
only one or two of the quasi-particles, one obtains the 
one- or two particle action given above, multiplied by
specific scattering factors
\item when the density operator acts on all three quasi-particles, more 
complicated many-body effects appear for $p\neq 1$.
One has to define a new function
\begin{equation}
G'_{ij}=\sum_{i=0}^{l-1} (-)^i \left(\begin{array}{c} m-l+i-1  \\ i \end{array} \right) \left(\begin{array}{c} l-1 \\ i \end{array} \right) \frac{\Gamma(g+i)\Gamma(m_2-m_1+g+1)\Gamma(m_2'-m_1'+g-i)}{\Gamma(g-i)\Gamma(m_2-m_1+g+i+1)\Gamma(m_2'-m_1'+g)}
\end{equation}
for two particles of the same type.
Here follow, case by case, the results we could obtain :
\begin{description}
\item[(2,1).] Acting on $|\{m_2,m_1,m_0=n_1\}>$, the factor 
multiplying $|\{m_2',m_1',m_0'\}>$ is
\begin{align}
&m\big[ \delta m_0 G_{01} G_{02} G'_{12} + \delta m_1 G_{01}G_{12} + \delta m_2 G_{02} G_{12} \nonumber \\ &+ (\delta m_0 \delta m_1 + \delta m_0 \delta m_2 + \delta m_1 \delta m_2 /p - \delta m_0 \delta m_1 \delta m_2 /(p-1) ) G_{01} G_{02} G_{12} \big]
\end{align}
with $\delta m_i = m_i-m_i'$ and the sum of the $\delta m_i$ being equal to $m$. Replacements similar to (\ref{eq:replacement}) have to be made, which is the case for any mixed state.
\item[(1,2).] Acting on $|\{n_2,n_1,n_0=m_1\}>$, the factor 
multiplying $|\{n_2',n_1',n_0'\}>$ is
\begin{align}
&m\big[ \delta n_0 G_{01} G_{02} G'_{12} / p(1-p) + \delta n_1 G_{01}G_{12} + \delta n_2 G_{02} G_{12} \nonumber \\ &+ (\delta n_1 \delta n_2 - \delta n_0 \delta n_1 \delta n_2 /(p-1) ) G_{01} G_{02} G_{12} \big] \ .
\end{align}
\item[(3,0).] Acting on $|\{m_3,m_2,m_1\}>$, the factor multiplying 
$|\{m_3',m_2',m_1'\}>$ is, for the special case $p$=2, 
\begin{align}
&-2m\big[ \delta m_1 G_{12} G_{13} + \delta m_2 G_{12}G_{23} + \delta m_3 G_{13} G_{23} \nonumber \\ &+ (\delta m_1 \delta m_2 -\delta m_2 \delta m_3 +3/2 \delta m_1 \delta m_2 \delta m_3 ) G_{12} G_{13} G_{23} \big] \ . \label{eq:3p}
\end{align}
\end{description}
\end{itemize}

In the next section the various form factors presented here will be 
used in a form factor expansion for the finite-temperature 
density-density correlation function.

\section{Form-factor expansion}
\label{sec:expansion}

We now turn to the goal of computing a finite-temperature correlation function.
Formally, it amounts to compute
\begin{equation}
\label{eq:corfun}
<\mathcal{O}>_T = \frac{1}{Z(T)} \sum_{\Psi\in\mathcal{H}_N} <\Psi|\mathcal{O}|\Psi> \exp(-\beta E_\Psi),
\end{equation}
where $\mathcal{H}_N$ is the Hilbert space of normalized states.
The problem arising with this expression is that it is hard to handle
for any given quantum field theory.
Indeed, divergences appear in the correlators of the right-hand side
and these need to be resummed. In the context of (massive) Integrable 
Field Theory (IFT), it has been proposed that, using the basis of the 
asymptotic particle states in the zero-temperature theory, one may be 
able to rewrite (\ref{eq:corfun}) as a single sum free of divergences
\cite{LeClairNPB552}. The resulting formula, called a 
`form factor expansion', can be evaluated by using scattering data,
the thermodynamic Bethe Ansatz (TBA) and the form factor bootstrap
(FFB) (see \cite{BalogNPB419} for a discussion of how the TBA
is recovered from the FFB). There is an ongoing discussion about
how precisely the form factor expansion can be implemented in IFTs, 
in particular for the case of multi-point correlation functions 
\cite{LeClairNPB552,DelfinoJPA34,MussardoJPA34,KonikXXX,SaleurNPB567}.

For the case of Conformal Field Theory, there has been a similar
but independent proposal for writing finite-temperature correlators
in terms of quasi-particle form factors and appropriate thermal
distribution functions \cite{vanElburgJPA33}. The idea here is
that the `fqH basis' (\ref{eq:fqHstates}) of eigenstates of the
CS hamiltonian provides the proper notion of `asymptotic particle 
states', with simple, but non-trivial fractional statistics 
properties.

The motivation for studying the form factor expansion for CFT has
been that, if successful, this approach to finite-temperature
correlation functions can possibly be extended to models, such
as quantum spin chains, that are gapless but that are not CFTs.

Comparing the proposed form factor expansions for IFT and CFT
(the particular CFT discussed in this paper), one is led to   
identify a 2-body $S$-matrix of the simple form
\begin{equation}
\boldsymbol{S}=\exp [ 2i\pi (\boldsymbol{\delta}-\boldsymbol{g})
\Theta (\theta) ]
\end{equation}
in the thermodynamic limit. It is still an open question whether 
the form factor computed in \cite{vanElburgJPA33} and in 
this paper can be obtained by means of an axiomatic approach,
starting from these scattering data.

In this paper, we study the following CFT form factor expansion for 
one-point functions
\begin{align}
<\mathcal{O}(\varepsilon)>_T &= \frac{1}{\beta} \sum_{M,N} O^{(M,N)} (\varepsilon) \\
O^{(M,N)} (\varepsilon)&= \beta a\sum_{\{m_i;n_j\}} D^{(M,N)}
(m,\{m_i;n_j\}) \prod_{i=1}^M \bar{n}_p (a m_i) \prod_{j=1}^N 
\bar{n}_{1/p} (a n_j) \\
D^{(M,N)}(m,\{m_i;n_j\}) &= {}_N<\{m_i;n_j\}| \mathcal{O}(m) |\{m_i;n_j\}>_N + \text{ subfactors} \ .
\label{eq:defirr}
\end{align}
Here $a=2\pi/L$, $\varepsilon=am$, $\mathcal{O}(\varepsilon)=a \mathcal{O}(m)$;
the continuum limit is obtained by sending $a\to 0$.
$D$ is the irreducible form-factor, which we shall define with precision.
In (\ref{eq:defirr}), the subfactors are built from multi-particle states which are subsets of $\{m_i;n_j\}$.
Their leading state of the form (\ref{eq:basis}) is the original one on which some creation operators are canceled.
It has then to be rewritten with the correct charge sector.
Specific examples of our definitions are,
\begin{align}
\label{eq:defirr-e}
D^{(2,0)}(m,\{m_2,m_1\}^Q) \equiv& {}_N^Q<\{m_2,m_1\}| \mathcal{O}(m) |\{m_2,m_1\}>_N^Q \nonumber \\ &- {}_N^Q<\{m_1\}| \mathcal{O}(m) |\{m_1\}>_N^Q - {}_N^Q<\{m_2+p\}| \mathcal{O}(m) |\{m_2+p\}>_N^Q,\\
D^{(0,2)}(m,\{n_2,n_1\}^Q) \equiv& {}_N^Q<\{n_2,n_1\}| \mathcal{O}(m) |\{n_2,n_1\}>_N^Q \nonumber \\ &- {}_N^Q<\{n_1\}| \mathcal{O}(m) |\{n_1\}>_N^Q - {}_N^{Q+1(-p)}<\{n_2(+1)\}| \mathcal{O}(m) |\{n_2(+1)\}>_N^{Q+1(-p)},
\label{eq:defirr-qh}
\end{align}
the charge shift $(-p)$ and the momentum shift $(+1)$ being present for $Q$=0.

This definition can be interpreted as follows.
Each form factor may appear as part of bigger form factors.
The way to resum them is to use these irreducible form factors and the one-particle distributions.
This expansion has already proved successful in describing the Green function for the $\nu=1/p$ fqH \cite{vanElburgJPA33}.
In the following we will concentrate on the density-density correlation function, where $\mathcal{O} (m)= p_m p_{-m}$. Most of the results will be given for any $p$, we will restrict to the $p=2$ case for numerics. 

Using standard CFT methods, one finds the density-density correlator to be
\begin{equation}
<\rho(-\varepsilon) \rho(\varepsilon)>_T = 
\frac{1}{\beta}\, \frac{p\, \beta\varepsilon}{e^{\beta \varepsilon}-1} \ .
\end{equation}
We shall compare the results of the form-factor expansion with this
exact result. We start (section A) by explaining how the procedure works 
for $p=1$. It already contains all the features and is completely solvable.
Then (sections B and C), we treat the general $p$ case, where we evaluate 
the asymptotics of the form factor expansion for $\beta \varepsilon\to$ 0 
or $\infty$ in closed form. Finally (section D), we discuss numerical
results for the full correlation function for the case $p=2$.

\subsection{Case $p$=1}

\begin{figure}
\begin{center}
\includegraphics{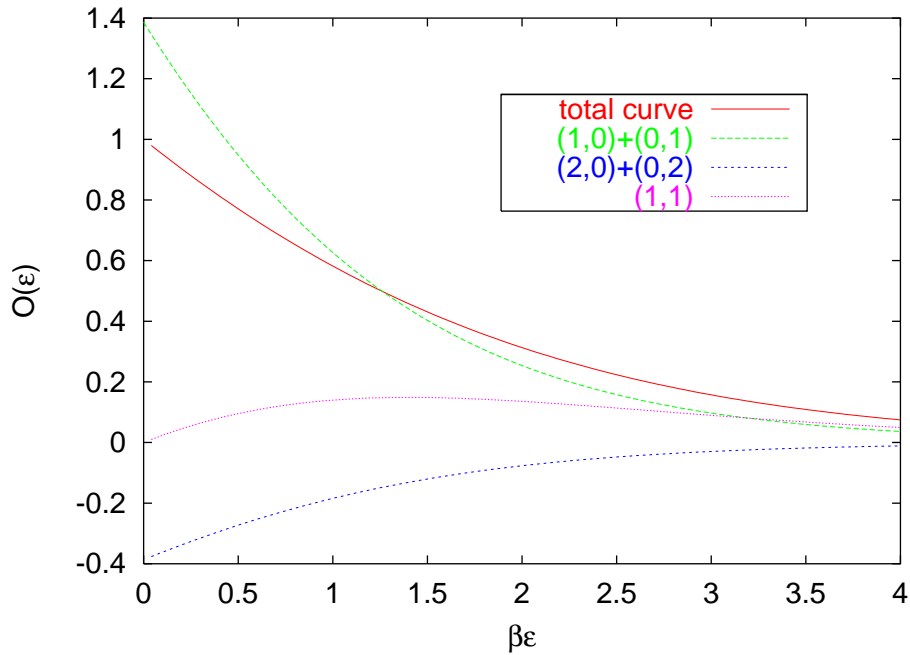}
\end{center}
\caption{\label{fig:Figp=1} All the form factor contributions at $p$=1.}
\end{figure}

The irreducible form factors are here straightforward to obtain.
One finds the following contributions to the density-density correlator
\begin{align}
O^{(1,0)}&=O^{(0,1)}=\beta\int_\varepsilon^\infty d\varepsilon_1\, \bar{n}_1(\varepsilon_1)=\log \left( \frac{e^{\beta\varepsilon}+1}{e^{\beta\varepsilon}} \right)\\
O^{(2,0)}&=O^{(0,2)}=-\beta\int_0^\infty d\varepsilon_1\, \bar{n}_1(\varepsilon_1)\bar{n}_1(\varepsilon+\varepsilon_1)=\frac{1}{e^{\beta\varepsilon}-1}\left[\log 2-e^{\beta\varepsilon} \log \left( \frac{e^{\beta\varepsilon}+1}{e^{\beta\varepsilon}} \right) \right]\\
O^{(1,1)}&=\beta\int_0^\varepsilon d\varepsilon_1\, \bar{n}_1(\varepsilon_1)\bar{n}_1(\varepsilon-\varepsilon_1)=\frac{2}{e^{\beta\varepsilon}-1}\log \left( \frac{e^{\beta\varepsilon}+1}{2e^{\beta\varepsilon/2}} \right) \ .
\end{align}
All the other irreducible form factor vanish.
One then easily checks that the sum of all contributions gives the correct answer $\beta\varepsilon/(\exp(\beta\varepsilon)-1)$.
The corresponding curves are shown in figure \ref{fig:Figp=1}.

Before going to general $p$, let us remark a few facts:
\begin{itemize}
\item the asymptotics for $\beta\varepsilon\to 0$ are
\begin{itemize}
\item $\beta\int_0^\infty d\varepsilon\, \bar{n}_1(\varepsilon)$ for 1 electron and 1 quasi-hole
\item $-\beta\int_0^\infty d\varepsilon\, \bar{n}_1^2(\varepsilon)$ 
for 2 electrons or 2 quasi-holes
\item 0 for 1 electron and 1 quasi-hole;
\end{itemize}
giving a total of $2\int_0^\infty d\beta\varepsilon\, \bar{n}_1(\varepsilon)(1-\bar{n}_1(\varepsilon))$=$2\bar{n}_1(0)$=1.
\item the asymptotics for $\beta\varepsilon\to\infty$ are
\begin{itemize}
\item $\beta e^{-\beta\varepsilon}$ for 1 electron or 1 quasi-hole
\item $\beta(\log 2-1)e^{-\beta\varepsilon}$ for 2 electrons or 2 quasi-holes
\item $\beta\varepsilon e^{-\beta\varepsilon}$ for 1 electron and 1 quasi-hole: it is the dominant one, and fits the exact asymptot.
\end{itemize}
\end{itemize}

These features for the asymptots will be shared for general $p$, as we will see now.

\subsection{Low-temperature expansion}

\begin{figure}
\begin{center}
\includegraphics{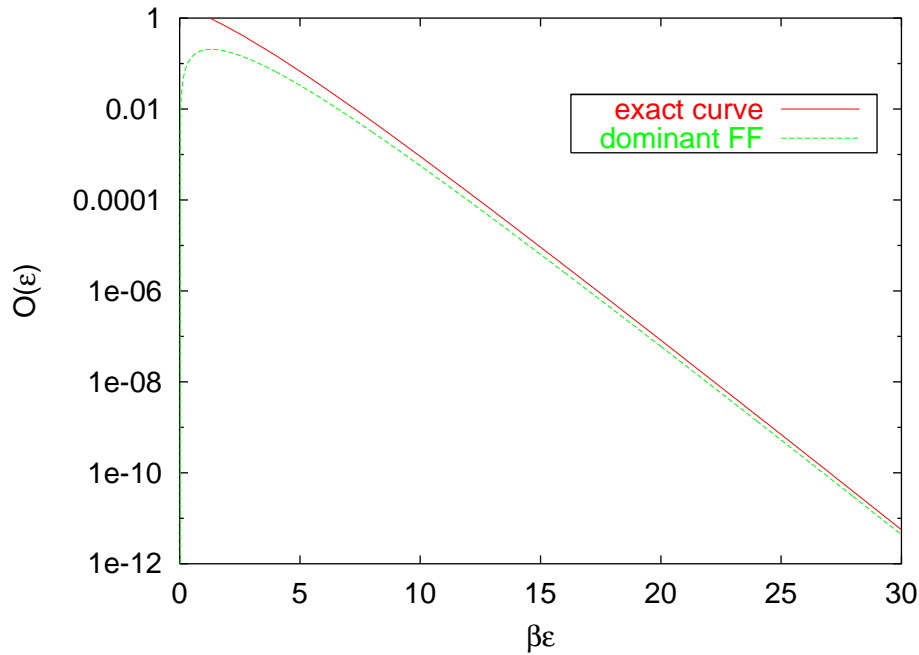}
\end{center}
\caption{\label{fig:low} Comparison of the exact $p=2$ low-temperature asymptot with the contribution from the form factor for a (1,2) state.}
\end{figure}

We look here at the $\varepsilon\to\infty$ limit, where the correlation 
function behaves as $p\beta\varepsilon e^{-\beta\varepsilon}$.

For a generic multi-particle excitation, this limit will be dominated by 
excitations for which one quasi-particle has a momentum $m_i$ bigger than 
$m=\varepsilon/a$, and all the others a much smaller one. The main
contribution comes from the term where the density operator acts only 
on the quasi-particle of momentum $m_i$. This contribution, which appears 
equally in all the subfactors of the irreducible form factor, is 
proportional to $\varepsilon^{1-g} \exp(-\beta\varepsilon)$.
The contribution from each other particle is of the form
\begin{align}
\text{contribution}(\varepsilon_{j\neq i}) &\propto \int_0^\infty d\varepsilon_j \bar{n}_{g_j}(\varepsilon_j) (S_{ij}-1)\\
&\propto \int_0^\infty d\varepsilon_j \bar{n}_{g_j}(\varepsilon_j) \frac{1}{(\varepsilon_i-\varepsilon)/\sqrt{g_i}\pm \varepsilon_j/\sqrt{g_j}}
\end{align}
where $S_{ij}$ is a scattering factor. These parts do not affect the leading 
behavior for $\varepsilon\to\infty$ and one finds a total contribution
which falls off faster than $\varepsilon e^{-\beta\varepsilon}$.

There is one exception to the above-mentioned rule.
The form factor can become big if the intermediate state is the vacuum, 
that is for an initial neutral excitation consisting of one electron and 
$p$ quasi-holes. To see this we use the expansion of $p_m$ in terms of 
Jack polynomials: we remark that for $\lambda=1/p$ this expansion is 
indeed restricted to states with 1 electron and $p$ quasi-holes.
The factor is then quickly found to be
\begin{equation}
D(1,2)(m=m_1+\sum_{j=1}^p n_j+p;m_1;\{n_j\}) = (\chi^{1/p}_\nu)^2 j^{1/p}_\nu,
\end{equation}
with $\{\nu\}=(\{n_j+1\},1^{m_1})$.
This expression has been evaluated in \cite{HaNPB435}. In the
thermodynamic limit $a\to 0$ it gives
\begin{multline}
O^{(1,p)}(\varepsilon) = \beta p^2\frac{\Gamma(p)\Gamma^p(1/p)}{\prod_{i=1}^{p}\Gamma^2(i/p)} \int d\varepsilon_1 \prod_{i=1}^{p} d\varepsilon_i' \delta(\varepsilon-\varepsilon_1 -\sum_i \varepsilon_i') \frac{\varepsilon^2\varepsilon_1^{p-1} \prod_{i<j} (\varepsilon_j'-\varepsilon_i')^{2/p}}{\prod_i(\varepsilon_1+p\varepsilon_i')^2 \prod_i (\varepsilon_i')^{1-1/p}} \\ \bar{n}_p(\varepsilon_1) \prod_i \bar{n}_{1/p}(\varepsilon_p') \text{  + subdominant terms} \ .
\end{multline}

Power counting in this formula gives an asymptotic behaviour proportional
to $\varepsilon e^{-\beta\varepsilon}$.
The coefficient in front is obtained by remarking that
\begin{equation}
O_T^{(1,p)}(\varepsilon) \simeq O_0^{(1,p)}(\varepsilon) e^{-\beta\varepsilon},
\end{equation}
For example, for $p=2$
\begin{align}
O^{(1,2)}(\varepsilon)
&\simeq \beta\varepsilon e^{-\beta \varepsilon} 4 \int d\varepsilon_1 d\varepsilon'_1 d\varepsilon'_2 \frac{\varepsilon \varepsilon_1 (\varepsilon'_2-\varepsilon'_1)}{\sqrt{\varepsilon'_1 \varepsilon'_2} (\varepsilon_1+2\varepsilon'_1)^2(\varepsilon_1+2\varepsilon'_2)^2} \Theta(\varepsilon'_2-\varepsilon'_1) \delta (\varepsilon-\varepsilon_1-\varepsilon'_1-\varepsilon'_2) \\
&\simeq \beta\varepsilon e^{-\beta \varepsilon} 4 \int_0^\infty d\theta \frac{\cosh (\theta)}{\sinh^3(\theta)} [\sinh(\theta)-\theta]\\
&=2 \beta\varepsilon e^{-\beta \varepsilon}
\end{align}
which is the right answer. The result is shown in figure \ref{fig:low}.

\subsection{High-temperature expansion}

When contemplating a form-factor expansion for finite temperature
correlators, one may worry about the convergence in the
high-temperature ($\beta\varepsilon\to 0$) limit, where thermal
distribution functions do not effectively suppress many-particle
contributions. We will see that in our case the situation
is remarkably good: using form factors with up to three 
quasi-particles, we recover the exact high-temperature limit.
To establish this result, we use the following identity
for the equilibrium distribution function $\bar{n}_g$
\cite{IsakovPRL83}
\begin{equation}
 \bar{n}_g(\varepsilon) 
 + (1-2g) \bar{n}_g^2 (\varepsilon) 
 - g(1-g) \bar{n}_g^3 (\varepsilon) 
 = - \bar{n}_g'(\varepsilon) \ .
\label{eq:ng-identity}
\end{equation}


\subsubsection{1 particle}

\begin{figure}
\begin{center}
\includegraphics{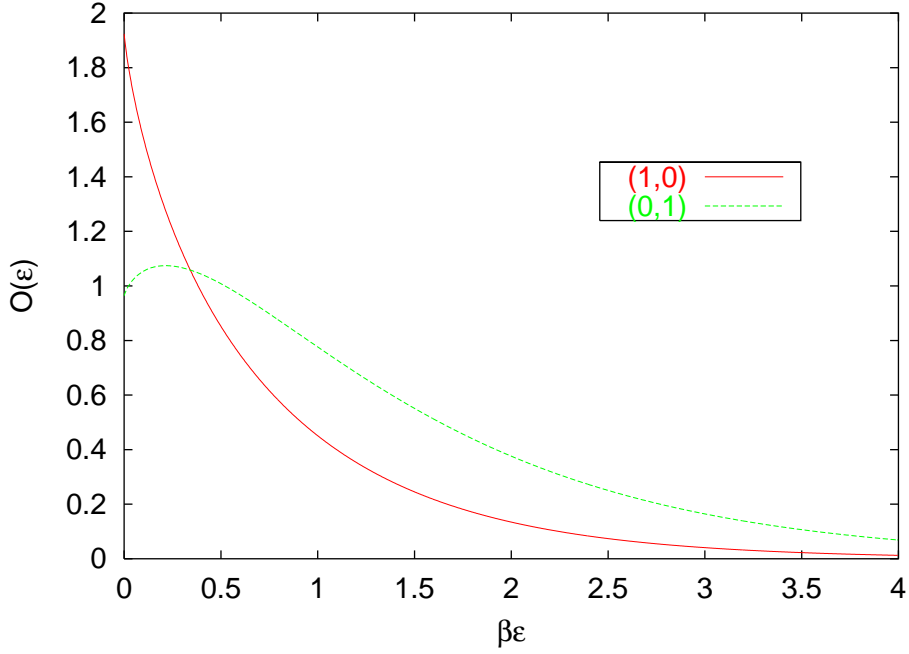}
\end{center}
\caption{\label{fig:1p} 1-particle contributions at $p$=2.}
\end{figure}

Direct application of (\ref{eq:ff1p}) gives the irreducible form factor
\begin{equation}
D^{(1)}(m;m_1) = pg \frac{\Gamma(m_1-m+g)\Gamma(m_1+1)}{\Gamma(m_1-m+1)\Gamma(m_1+g)} \ .
\end{equation}
The form factor contribution is then
\begin{equation}
O^{(1)}(\varepsilon\to 0)=pg \, 
\beta\int d\varepsilon_1\, \bar{n}_g(\varepsilon_1).
\end{equation}
On figure \ref{fig:1p} are the curves obtained at $p=2$ for one electron and one quasi-hole.

\subsubsection{2 particles}

\begin{figure}
\begin{center}
\includegraphics{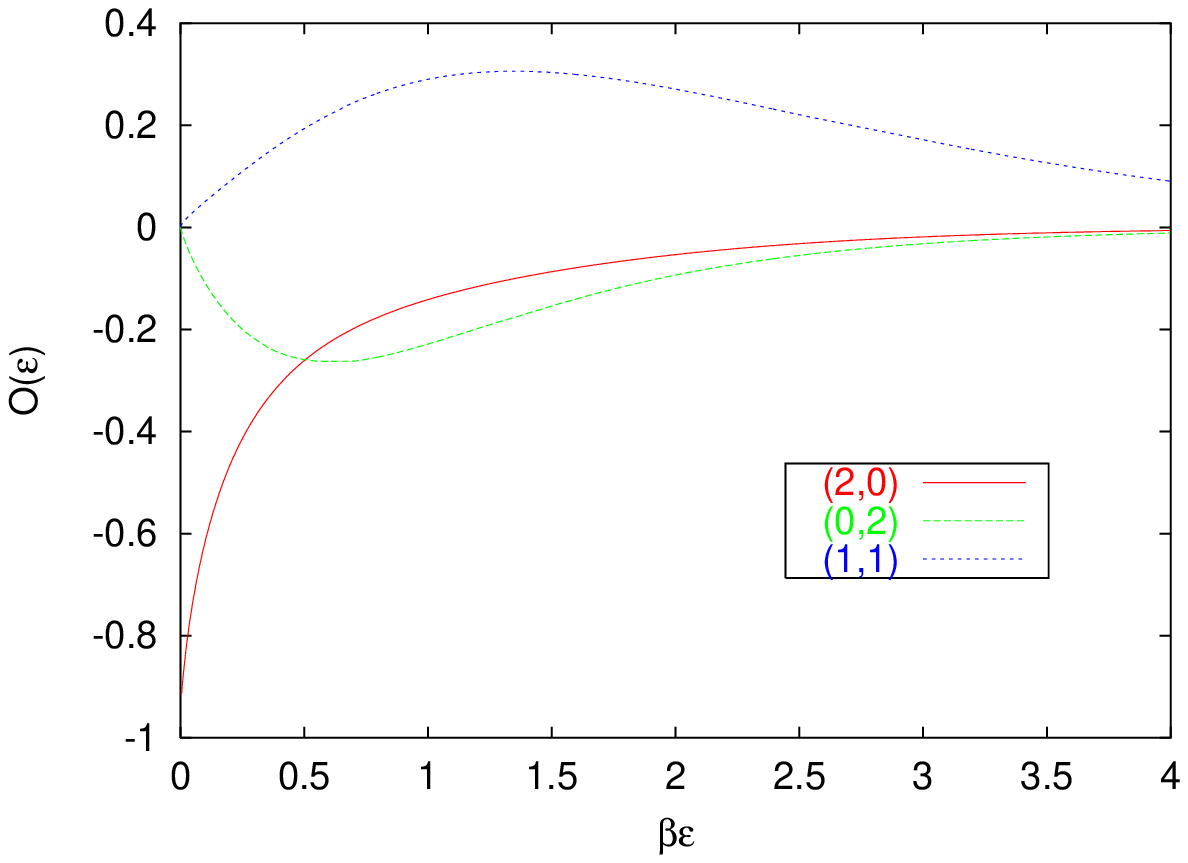}
\end{center}
\caption{\label{fig:2p} 2-particle contributions at $p$=2.}
\end{figure}

Using (\ref{eq:ff2p}) and the definitions for the irreducible form factor (\ref{eq:defirr-e},\ref{eq:defirr-qh}), one obtains for $(m_1,m_2) \gg m$ (2 particles of the same type)
\begin{align}
D^{(2)}(m;m_2,m_1)&=pg2\sum_{i=1}^m (-)^i \left(\begin{array}{c} m-1 \\ i-1 \end{array} \right) \left(\begin{array}{c} m+i-1 \\ i \end{array} \right) \frac{\Gamma(g+i)\Gamma(m_2-m_1+g-i)\Gamma(m_2-m_1+g+1)}{\Gamma(g-i)\Gamma(m_2-m_1+g)\Gamma(m_2-m_1+g+i)} \\
&\simeq pg\delta_{m_2,m_1}(2g-1) \sum_{i=1}^m \frac{(-)^i}{2i-1} \left(\begin{array}{c} m-1 \\ i-1 \end{array} \right) \left(\begin{array}{c} m+i-1 \\ i \end{array} \right)\\
&\simeq -pg(2g-1) \delta_{m_2,m_1} \ ,
\end{align}
leading to
\begin{equation}
O^{(2)}(\varepsilon\to 0)=
-pg(2g-1)\, \beta\int d\varepsilon_1\, \bar{n}^2_g(\varepsilon_1) \ .
\end{equation}

The curves for two electrons and two quasi-holes at $p=2$ are given in figure \ref{fig:2p}. As can be seen, the limit $\varepsilon \to 0$ is indeed 
$-12\,\beta\int d\varepsilon_1 \, \bar{n}_2^2(\varepsilon_1)=-0.9472$ 
for 2 electrons, and 0 for 2 quasi-holes.

For mixed states (1 electron and 1 quasi-hole), (\ref{eq:ffmix}) leads to the following limit for the irreducible form factor
\begin{equation}
O^{(1,1)}(\varepsilon\to 0) \sim (p^2+1) \beta\varepsilon \int_\varepsilon d\varepsilon_1 d\varepsilon_1' \frac{1}{(\varepsilon_1+p\varepsilon_1')^2} \bar{n}_p(\varepsilon_1) \bar{n}_{1/p}(\varepsilon_1'),
\end{equation}
leading to a linear in $\varepsilon$ dependence. 
So it has a vanishing limit for $\varepsilon\to 0$.
This result will be the same for any mixed state.

\subsubsection{3 particles}

\begin{figure}
\begin{center}
\includegraphics{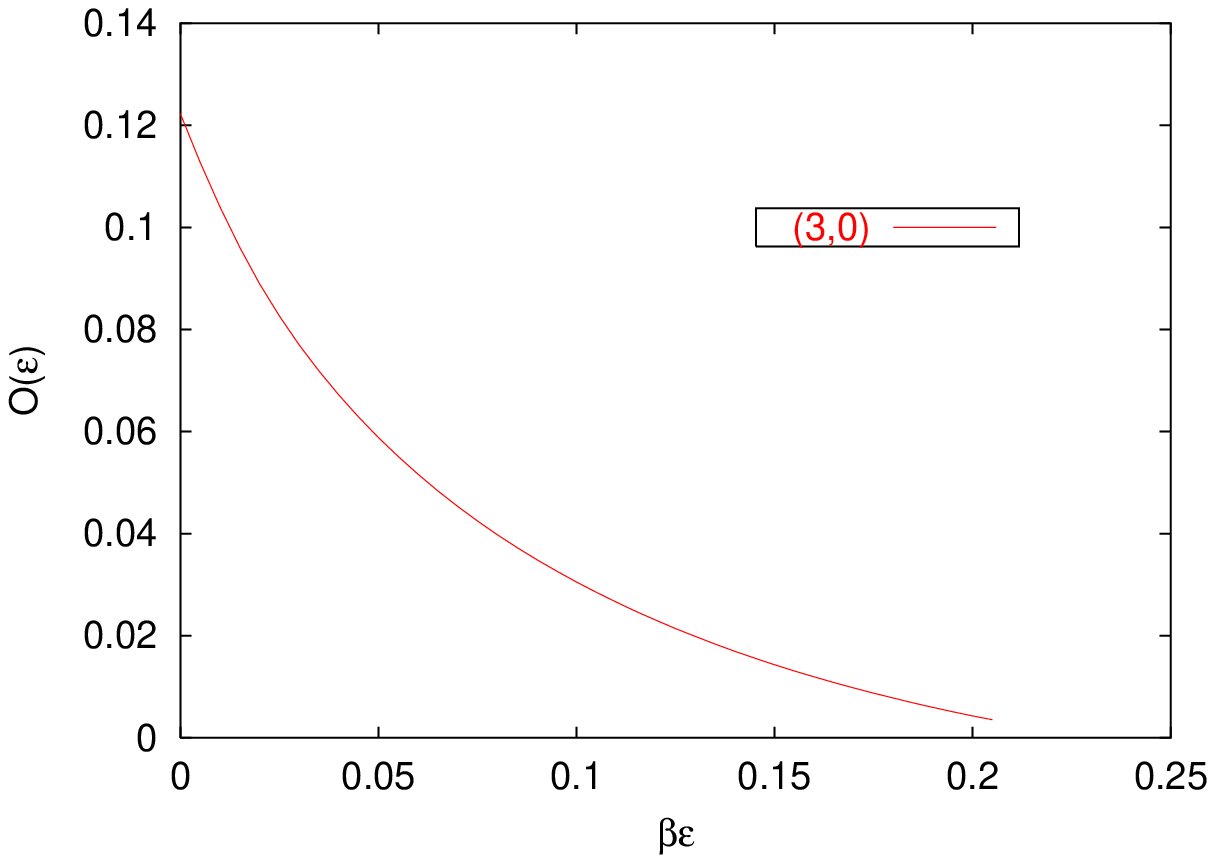}
\end{center}
\caption{\label{fig:3p} 3-particle contribution at $p=2$ in the 
high-temperature limit. Due to the complexity of the numerical 
calculations, the discretization used for $\beta\varepsilon$ is 
$5\times10^{-3}$, a relatively large value which accounts for the 
under-estimation of the $\varepsilon\to 0$ limit.}
\end{figure}

Considering the argument given above, we are only interested in the 
contribution of 3 particles of the same type (3 electrons
or 3 quasi-holes). As for 2 particles, we find that the contribution 
in the limit $\beta\varepsilon\to 0$ comes entirely from the diagonal, 
where the energies of the three particles are the same. We have not been
able to get an analytic expression for the high-temperature limit,
but we present the following conjecture
\begin{equation}
\label{eq:irr3p}
O^{(3)} (\varepsilon\to 0) = pg^2(g-1) \,
\beta\int d\varepsilon_1\, \bar{n}_g^3(\varepsilon_1) \ .
\end{equation}

We have strong support for this conjecture. First of all, thanks to 
(\ref{eq:3p}), we are able to compute the contributions from form 
factors for 3 electrons at $p=2$ for low values of $m>0$, all
representing the limit $\varepsilon=am\to 0$. We reproduce the 
expected form (\ref{eq:irr3p}) when $m=3$, and we expect that
this result is stable for $m\geq 3$. Continuity for bigger 
values of $m$ (corresponding to finite $\varepsilon=am$) has 
been checked numerically, as is shown in figure \ref{fig:3p}. 
[We note that according to eq.~(\ref{eq:irr3p}) the numerical 
value of the $(3,0)$ contribution at $\varepsilon=0$ will be 
.1272; the numerical curve displayed in figure \ref{fig:3p} 
gives a slightly smaller value, due to the fact that a 
relatively large value of the discretization $a$ had to be used.] 
Further support for the conjecture comes from the irreducible 
form factor for 3 quasi-holes for general $p$ at $m$=1, which 
directly gives the expression (\ref{eq:irr3p}).

\subsubsection{Sum of 1, 2 and 3 particle contributions}

We are now ready to evaluate the sum of all contributions to
the density-density correlator at $\beta\varepsilon\to 0$.
Adding the contributions from form factors
with 1, 2, or 3 quasi-particles of statistics $g$ results in
\begin{equation}
 pg \beta\int_0^\infty d\varepsilon
\left( \bar{n}_g(\varepsilon) 
       + (1-2g) \bar{n}_g^2 (\varepsilon) 
       -g(1-g) \bar{n}_g^3 (\varepsilon) \right) 
  = - pg \beta\int_0^\infty d\varepsilon\, \bar{n}_g'(\varepsilon) 
  = pg \bar{n}_g (0) .
\label{eq:distr-id}
\end{equation}
The density-density correlator has these contributions
from electrons ($g=p$) and from quasi-holes ($g=1/p$). Through 
duality $g\bar{n}_g(0)+1/g\, \bar{n}_{1/g}(0) = 1$, leading to 
the result 
\begin{equation}
\beta \, <\rho(-\varepsilon) \rho(\varepsilon)>_T 
\, \stackrel{\beta\varepsilon\to 0}{\to} \, p
\end{equation}
in agreement with the exact result. We thus find that the high
temperature limit is saturated by contributions with up
to three quasi-particles, with the 3-particle contributions
being absent in the special case $p=1$. In our view, this
non-trivial, exact result gives strong support for the validity 
of the proposed CFT form factor expansion. We mention that,
in general, good convergence at the high-temperature end is 
scarcely seen in form factor expansions.

\subsection{Full correlation function}

\begin{figure}
\begin{center}
\includegraphics{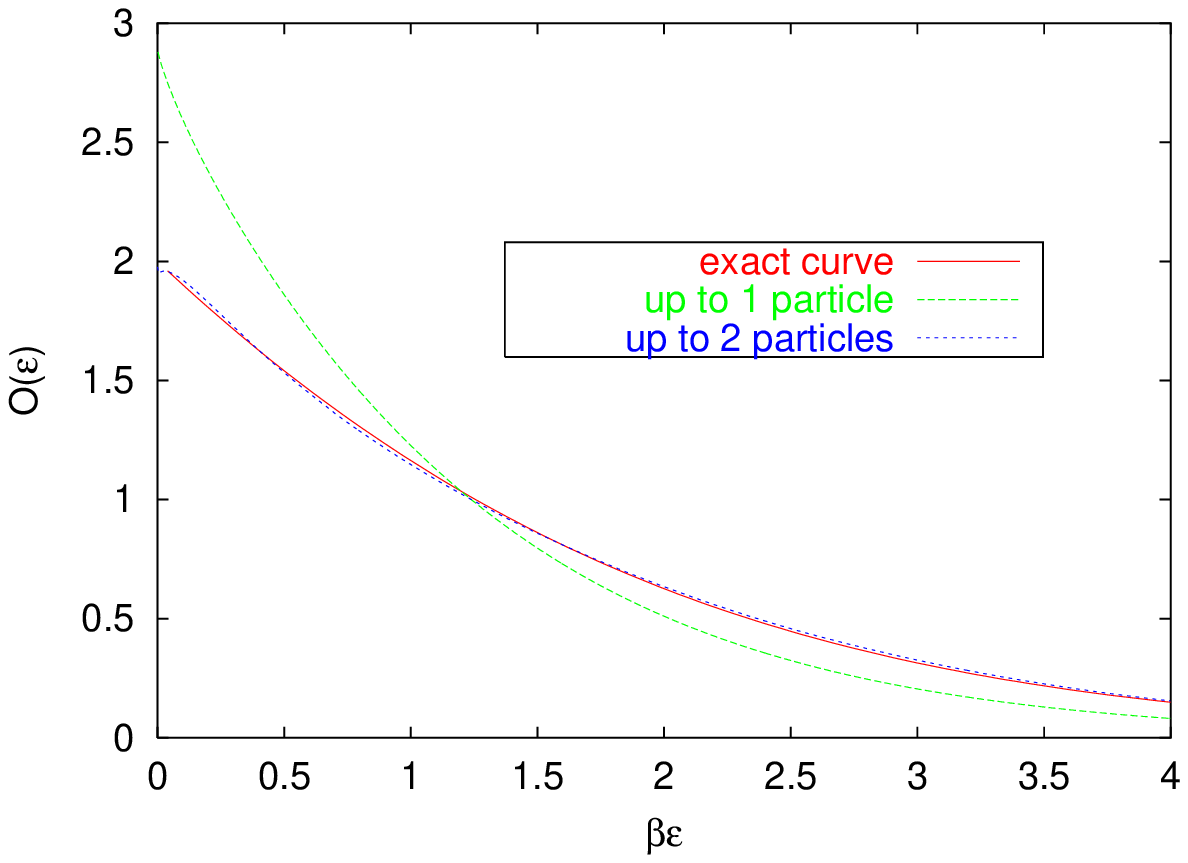}
\end{center}
\caption{\label{fig:tot} High temperature expansion. One can compare the exact density-density correlation function at $p=2$ and the contributions from up to 1 and 2 quasi-particles. The agreement is very good in the latter case.}
\end{figure}

In the two preceding sections, we have proved that the low-
and high-temperature asymptotics for the density-density correlation 
function were recovered in the form factor expansion using only a 
few quasi-particles. We now observe that the integrated value
of the correlator is recovered already at the level of the 1 particle 
contributions. One only needs the sum rule (\ref{eq:sumrule}) and the 
procedure for building irreducible form factors to prove this. At the 
level of 1 particle, one gets
\begin{equation}
\beta \int_0^\infty d\varepsilon\, 
O^{(1)} = p \beta \int_0^\infty d\varepsilon\, \, 
\beta\varepsilon(\bar{n}_p + \bar{n}_{1/p})(\varepsilon) 
= p \beta \int_0^\infty d\varepsilon\, 
\frac{\beta\varepsilon}{e^{\beta \varepsilon}-1} \ .
\end{equation}
At the level of 2 particles,
\begin{equation}
\beta \int_0^\infty d\varepsilon\, O^{(2)}= -\frac{1}{2} \big[ \beta 
\int_0^\infty 
d\varepsilon\, (p \bar{n}_p-\bar{n}_{1/p})(\varepsilon)\big]^2 = 0
\end{equation}
by duality.
For any contribution with more than 2 particles, the integrated curve is 0.

All these facts tend to show that the proposed form factor expansion gives 
a quick convergence for the \emph{full} correlation function.
This is indeed the case, as one sees on figure \ref{fig:tot},
where the contribution from states with up to 2 particles at 
$p$=2 have been collected.

\section{Conclusion}

We have developed a form factor expansion for the density-density 
correlation function in the fqH edge CFT at $\nu=1/p$.
We have used the correspondence between a fqH basis made of dual 
quasi-particles (electrons and quasi-holes) and the Jack polynomial basis 
of the Calogero-Sutherland model. We used `Jack polynomial technology' 
to study form factors, and we obtained analytical expressions for form 
factors with up to 3 excited particles. Using these form factors in the 
proposed form factor expansion, we showed that the convergence towards 
the exact correlation function is quick in terms of the number of excited 
particles. We demonstrated that in the high- and low- temperature regimes, 
the expansion reproduces the exact result. This is a clear indication 
that the expansion is correct.

This result opens the way for the computation of correlation functions for 
non-CFT theories, as soon as their excitations form a gas of fractional 
statistics particles.
This is the case for the Calogero-Sutherland model, and the models derived 
from it, like the Haldane-Shastry spin chain.
Of fundamental interest also is the need to understand the connection 
between the thermodynamic limit of the CFT form factors and a Thermodynamic 
Bethe Ansatz method. This would allow for treating such complicated problems 
as impurities in edge-to-edge tunneling.

\vskip 2mm

We thank Robert Konik for discussions and comments on the 
manuscript. This research was supported in part by the 
Netherlands Organisation for Scientific Research (NWO). 
KS was supported by the NSF under grant no. DMR-98-02813.

\appendix*
\section{Jack polynomial technology}

Most of the Jack polynomial technology is found in \cite{Macdonald}. 
We reproduce here the results useful in the present paper.

Jack polynomials form a complete and orthogonal basis of the ring of symmetric polynomials with the given inner product:
\begin{equation}
<p_\mu, p_\nu>=\delta_{\mu,\nu} \lambda^{-l(\mu)} z_\mu
\end{equation}
where $p_\mu = \prod_{j=1}^{l(\mu)} p_{\mu_j}$, $p_m(\{x_i\})=\sum_i x_i^m$ are the power sums, and $z_\mu = \prod_{i >= 1} i^{m_i}m_i!$.

They are defined in a unique way through their development in monomial symmetric functions:
\begin{equation}
P^\lambda_\mu = \sum_{\nu <= \mu} v_{\mu,\nu}^\lambda m_\nu \qquad \text{with} \quad v_{\mu,\mu}=1.
\end{equation}

Their norms are
\begin{equation}
<P^\lambda_\mu,P^\lambda_\mu> = j^{\lambda}_{\{\nu\}}=\prod_{s\in\{\nu\}} \frac{\lambda l(s) +a(s)+1}{\lambda(l(s)+1)+a(s)},
\label{eq:Jackinner}
\end{equation}
where $l(s)$ and $a(s)$ are the leg and the arm of cell $s$ in the tableau $\{\nu\}$.

In the paper, we need the expansion of power sums in Jack polynomials:
\begin{align}
\label{eq:pm/Jack}
p_m &= \sum_{\mu\vdash m} \chi^\lambda_\mu J^\lambda_\mu \\
\chi^\lambda_\mu &= m \frac{\prod_{s\neq(1,1)} (a'(s)-\lambda l'(s))}{\prod_s (\lambda l(s)+a(s)+1)}
\end{align}
and in elementary polynomials:
\begin{equation}
\label{eq:pm/e}
p_m=m\sum_{\nu\vdash m}\, (-1)^{m-l(\nu)} \, \frac{(l(\nu)-1)!}{\prod_i m_i(\nu)!} e_\nu,
\end{equation}
where $e_\nu=e_{\nu_1} \ldots e_{\nu_m}$ and $e_r=J^\lambda_{(1^r)}$ for any $\lambda$. For example,
\begin{align}
p_1 &= e_1 \\
p_2 &= e_{1^2} - 2e_2 \\
p_3 &= e_{1^3} - 3e_{21} + 3e_3  \\
p_4 &= e_{1^4} - 4e_{21^2} + 2e_{2^2} + 4e_{31} - 4e_4 \\
p_5 &= e_{1^5} - 5e_{21^3} + 5e_{2^21} + 5e_{31^2} -5e_{32} -5e_{41} +5e_5 \\
p_6 &= e_{1^6} - 6e_{21^4} + 9e_{2^21^2} - 2e_{2^3} + 6e_{31^3} - 12e_{321} + 3e_{3^2} - 6e_{41^2} + 6e_{42} + 6e_{51} - 6e_6 \ .
\end{align}

Inner products between Jacks and elementary symmetric functions are known through the Pieri formula:
\begin{align}
\label{eq:Pieri}
J^\lambda_\nu e_r &=  \sum_\mu \psi'_{\mu/\nu} J^\lambda_\mu,\\
\psi'_{\mu/\nu} &=  \prod_{C_{\mu/\nu}\backslash R_{\mu/\nu}} \frac{j^\lambda_\mu(s)}{j^\lambda_\nu(s)}
\end{align}
with $\mu-\nu$ being a vertical $r$-strip (at maximum 1 box per row, for a total of $r$), $C_{\mu/\nu}$ (resp. $R_{\mu/\nu}$) being the union of columns (resp. rows) that intersect $\mu-\nu$.

\bibliography{density}

\end{document}